\documentclass[twocolumn,superscriptaddress,showpacs,preprintnumbers,amsmath,amssymb,oneside,prl]{revtex4}
\usepackage{graphicx}
\usepackage{dcolumn}
\usepackage{bm}

\begin{document}


\title{Generation of Narrow-Band Polarization-Entangled Photon Pairs for Atomic Quantum Memories}

\author{Xiao-Hui Bao}
\affiliation{Hefei National Laboratory for Physical Sciences at
Microscale and Department of Modern Physics, University of Science
and Technology of China, Hefei, Anhui 230026, China}
\affiliation{Physikalisches Institut der Universitaet Heidelberg,
Philosophenweg 12, Heidelberg 69120, Germany}

\author{Yong Qian}
\affiliation{Hefei National Laboratory for Physical Sciences at
Microscale and Department of Modern Physics, University of Science
and Technology of China, Hefei, Anhui 230026, China}

\author{Jian Yang}
\affiliation{Hefei National Laboratory for Physical Sciences at
Microscale and Department of Modern Physics, University of Science
and Technology of China, Hefei, Anhui 230026, China}

\author{Han Zhang}
\affiliation{Hefei National Laboratory for Physical Sciences at
Microscale and Department of Modern Physics, University of Science
and Technology of China, Hefei, Anhui 230026, China}

\author{Zeng-Bing Chen}
\affiliation{Hefei National Laboratory for Physical Sciences at
Microscale and Department of Modern Physics, University of Science
and Technology of China, Hefei, Anhui 230026, China}

\author{Tao Yang}
\affiliation{Hefei National Laboratory for Physical Sciences at
Microscale and Department of Modern Physics, University of Science
and Technology of China, Hefei, Anhui 230026, China}

\author{Jian-Wei Pan}
\affiliation{Hefei National Laboratory for Physical Sciences at
Microscale and Department of Modern Physics, University of Science
and Technology of China, Hefei, Anhui 230026, China}
\affiliation{Physikalisches Institut der Universitaet Heidelberg,
Philosophenweg 12, Heidelberg 69120, Germany}

\date{\today}


\begin{abstract}
We report an experimental realization of a narrow-band
polarization-entangled photon source with a linewidth of 9.6 MHz
through cavity-enhanced spontaneous parametric down-conversion. This
linewidth is comparable to the typical linewidth of atomic ensemble
based quantum memories. Single-mode output is realized by setting a
reasonable cavity length difference between different polarizations,
using of temperature controlled etalons and actively stabilizing the
cavity. The entangled property is characterized with quantum state
tomography, giving a fidelity of $94\%$ between our state and a
maximally entangled state. The coherence length is directly measured
to be 32 m through two-photon interference.
\end{abstract}


\maketitle

The storage of photonic entanglement with quantum memories plays an
essential role in linear optical quantum computation (LOQC)
\cite{klm2001} to efficiently generate large cluster states
\cite{Browne2005}, and in long-distance quantum communication (LDQC)
to make efficient entanglement connections between different
segments in a quantum repeater \cite{Briegel1998}. For the atomic
ensemble based quantum memories \cite{Chaneliere2005, Choi2008,
Chen2008}, typical spectrum linewidth required for photons is on the
order of several MHz. While spontaneous parametric down-conversion
(SPDC) is the main method to generate entangled photons
\cite{kwiat1995}, the linewidth determined by the phase-matching
condition is usually on the order of several THz which is about
$10^6$ times larger, making it unfeasible to be stored. Moreover,
interference of independent broad-band SPDC sources requires a
synchronization precision of several hundred fs \cite{Zukowski1995}.
While in LDQC, for the distance on the order of several hundred km,
it becomes extremely challenging for the current synchronization
technology \cite{Kaltenbaek2006, Yang2006}. But for a narrow-band
continuous-wave source at MHz level, due to the long coherence time,
synchronization technique will be unnecessary, while coincidence
measurements with time resolution of several ns with current
commercial single-photon detectors will be enough to interfere
independent sources.

Passive filtering with optical etalons is a direct way to get MHz
level narrow-band entangled photons from the broad-band SPDC source,
but it will inevitably result in a rather low count rate. In
contrast, cavity-enhanced SPDC \cite{Ou1999,Shapiro2000} provides a
good solution for this problem. By putting the nonlinear crystal
inside a cavity, the generation probability for the down-converted
photons whose frequency matches the cavity mode will be enhanced
greatly. The cavity acts as an active filter. The frequency of the
generated photons lies within the cavity mode, which can be easily
set to match the required atomic linewidth. Experimentally, Ou
\textit{et al.} \cite{Ou1999} has realized a type-I source, in which
the two photons generated have the same polarization, making it very
difficult to generate entanglement. Wang \textit{et al.}
\cite{Wang2004} made a further step by putting two type-I nonlinear
crystals within a ring cavity to generate polarization entanglement,
but unfortunately the output is multi-mode which does not fit the
requirement of an atomic quantum memory. While, a type-II configured
source (down-converted photons have different polarizations) is more
preferable for the ease of generating polarization entanglement,
compared with a type-I source. Recently Kuklewicz \textit{et al.}
\cite{Kuklewicz2006} have realized a type-II source, but the output
is still multi-mode. So far to the best of our knowledge, a true
narrow-band (single-mode) polarization-entangled photon source at
MHz level has never been reported yet along this line. Direct
generation of narrow-band photon pairs from cold atomic ensembles
\cite{Thompson2006, Yuan2007} is another solution, but the setup is
usually much more complicated.

In this Letter, we show that for a type-II cavity-enhanced source by
setting a cavity length difference between different polarizations
we can reduce the proportion for the background modes greatly.
Further suppression is realized by using filter etalons to make the
output single-mode. Entanglement is generated by interfering the
photon pair on a polarizing beam-splitter (PBS). The measured
linewidth for this narrow-band polarization-entangled source is 9.6
MHz, which is comparable to the typical linewidth of atomic quantum
memories.

\begin{figure}[hbtp]
\includegraphics[width=1\columnwidth]{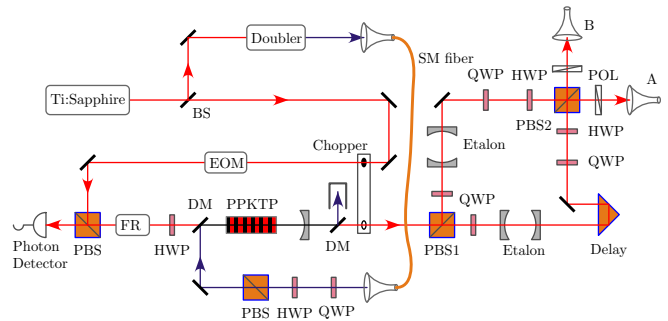}
\caption{(color online). Layout of the experiment. The generation of
polarization entanglement is realized by interfering the two
down-converted photons on PBS2. An electro-optic modulator (EOM)
phase modulated at 70 MHz is utilized to generate sidebands for the
locking beam. A PBS, a Faraday rotator (FR) and a half-wave plate
(HWP) is utilized to extract the reflection beam and to generate the
error signal for the locking system.}\label{setup}
\end{figure}

In our experiment, a flux-grown periodically poled KTiOPO4 (PPKTP)
crystal (1cm long) is used as the nonlinear medium. Quasi-phase
matching is optimized for a horizontally (H) polarized ultraviolet
(UV) pump photon (390 nm) down-converting to a near-infrared photon
pair (780 nm) with one polarized in H and the other in vertical (V).
The phase-matching bandwidth is 175 GHz. The first side of the PPKTP
is high-reflection coated (R $>$ 99\% at 780 nm) to form the
double-resonant cavity with a concave mirror (R $\approx$ 97\% at
780 nm) of 10-cm curvature, as shown in FIG. \ref{setup}. The second
side of the PPKTP is anti-reflection (AR) coated to minimize losses
within the cavity. Both the PPKTP and the concave mirror are AR
coated at 390 nm so that the UV pump interacts only once with the
PPKTP in the cavity.

The cavity is intermittently locked using the Pound-Drever-Hall
scheme \cite{Black2001}. A mechanical chopper is designed to block
the cavity output when the locking beam is switched on, to avoid the
leaking beam entering into posterior single-photon detectors. This
locking system is only effective for the cavity noise whose
frequency is much lower than the locking repetition rate (50 Hz),
i.e. the long-term drift. In order to suppress the high frequency
noise, especially the strong acoustic noise at subkilohertz, we
build the cavity from a single block of stainless steel by digging
out the inner part. The PPKTP crystal along with the oven, the
thermal electric cooler (TEC) and the concave mirror are fixed
firmly inside. The steel block is covered from lateral side with two
pieces of organic glass to prevent airflow. Temperature of the PPKTP
crystal is controlled to the precision of about 0.002 $^\circ
\rm{C}$ with a high-performance temperature controller. The
frequency of the locking beam is the same as the center frequency
($\omega_0$) of the down-converted photons. Since the polarization
of the locking beam is rotated to H before entering the cavity, this
active locking system can only guarantee the resonance at $\omega_0$
for H. The resonance of the cavity at $\omega_0$ for V is realized
by slightly tuning the temperature of the PPKTP.

It has been pointed out that for the case of type-II configured
cavity-enhanced SPDC, ideally the output will be single-mode
\cite{Lu2000}. But considering into the finite finesse of the
cavity, the ideal single-mode output will be mixed with several
nearby background modes. The quantum state can be expressed as:
\begin{eqnarray}
|\Psi\rangle&=&\sqrt{\chi_0}|\omega_0\rangle_H|\omega_0\rangle_V\nonumber\\
&+&\sum_{m=1}^{N=46}
\frac{\sqrt{\chi_m}}{2}(|\omega_0+m\Omega_H\rangle_H|\omega_0-m\Omega_H\rangle_V\nonumber\\
&&+|\omega_0-m\Omega_H\rangle_H|\omega_0+m\Omega_H\rangle_V\nonumber\\
&&+|\omega_0+m\Omega_V\rangle_H|\omega_0-m\Omega_V\rangle_V\nonumber\\
&&+|\omega_0-m\Omega_V\rangle_H|\omega_0+m\Omega_V\rangle_V)
\end{eqnarray}
with
\begin{equation}
\frac{\chi_m}{\chi_0}=\frac{4}{1+\frac{4F^2}{\pi^2}\sin^2{\frac{m\Delta\Omega}{\Omega}\pi}}
\end{equation}
where $\Omega_H$ and $\Omega_V$ are the free spectrum ranges (FSR)
for H and V respectively, with the average value of $\Omega$ (1.9
GHz) and the difference of $\Delta \Omega$ (21 MHz); $F$ is the
finesse of the cavity, with the measure value of 166; $N$ is determined by the
phase-matching bandwidth of PPKTP. The first term of the right side
of Eq. (1) is the expected single-mode output. The following four
terms in the summation correspond to the case that one photon is
resonant with the cavity while the other is not. For the modes near
$\omega_0$, we have $\chi_1/\chi_0\,=\,1.7$,
$\chi_2/\chi_0\,=\,0.63$, $\chi_3/\chi_0\,=\,0.31$. When $m$ goes
higher, $\chi_m$ asymptotically goes to 0. While in the case of
equal cavity length ($\Delta \Omega$ = 0) \cite{Kuklewicz2006}, each
$\chi_m$ nearly has the same value. The ratio between the summation
of these backgound modes to the center mode is 3.41. In our
experiment, we use etalons (FSR\,=\,13.9\,GHz, Finesse\,=\,31) to
eliminate these nearby modes. The etalons are put into separate
copper ovens, and temperature controlled to the precision of 0.01
$^\circ \rm{C}$ to achieve a stable performance.

The complete experimental setup is illustratively shown in FIG.
\ref{setup}. A frequency-stabilized Ti:Sapphire laser with a
linewidth of 75 kHz is utilized as the main laser. A small
proportion of the output power is split as the locking beam, and the
rest power is sent to an eternal-cavity frequency doubler. The
generated UV beam is very elliptical, and we use several cylindrical
lenses to convert it to a near-Gaussian beam, which is further
coupled into a single-mode fiber and released later with a fiber
collimator. Two high-performance dichroic mirrors (DM) are used to
combine the UV pump with the locking beam, and later separate the
remained UV pump from the cavity output. The generated photon pair
is separated on PBS1, and filtered with separated etalons. In our
experiment we find that slight reflection from the etalons will
cause the double-resonant cavity very unstable. We add a
quarter-wave plate (QWP) in each path to form a optical isolator
with PBS1 to eliminate the etalon reflection.

\begin{figure}[hbtp]
\includegraphics[width=.8\columnwidth]{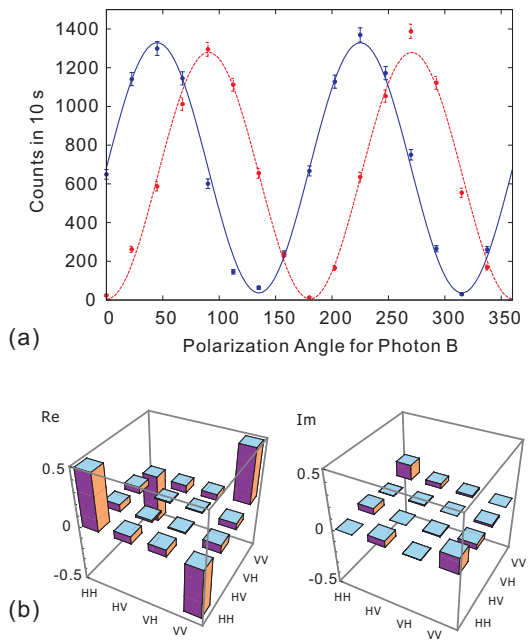}
\caption{(color online). (a) Polarization correlations for the
entangled photon pair. Polarization angle for photon A is fixed to
$90^\circ$ for the red, and $-45^\circ$ for the blue. Error bars
represent statistical errors. (b) Tomography measurement result,
with the left for the real part, the right for the imaginary part.}
\label{fig2}
\end{figure}

By making a two-photon interference for the narrow-band photons on
PBS2, with one photon polarized in $|+\rangle = 1 / \sqrt{2}
(|H\rangle + |V\rangle)$  and the other in $|-\rangle = 1 / \sqrt{2}
(|H\rangle - |V\rangle)$ as input, we are able to generate
polarization entanglement for the case one photon in each output
port. These two photons are further coupled into single-mode fibers
and detected with single-photon detectors.The desired output state
is
$|\phi^-\rangle=1/\sqrt{2}(|H\rangle|H\rangle-|V\rangle|V\rangle)$.
But during the overlapping on PBS2 there is some phase shift between
$|H\rangle$ and $|V\rangle$, leading to an output state of
$1/\sqrt{2}(|H\rangle|H\rangle-e^{i\alpha}|V\rangle|V\rangle)$. We
insert an adjustable wave-plate in one output path to compensate
this phase shift. In order to verify the entanglement property, we
first make a polarization correlation measurement at 4 mW pump
power, with the result shown in FIG.\ref{fig2}(a). The visibility is
about $97\%$, which is far beyond the requirement for a violation of
Bell-CHSH inequality \cite{chsh1969}. In this inequality the value
$S$ is defined as
\begin{equation}
S=|E(\phi_A,\phi_B)-E(\phi_A,\phi_B')+E(\phi_A',\phi_B)+E(\phi_A',\phi_B')|
\end{equation}
where $\phi_A$ and $\phi_A'$ are two polarization measuring angles
for photon A, and $\phi_B$ and $\phi_B'$ for photon B, and
$E(\alpha,\beta)$ is the correlation coefficient between these two
photons. Violation of this inequality ($S > 2$) is a direct proof of
entanglement. Our measured result is $S = 2.66 \pm 0.03$, for which
the inequality is violated by 22 standard deviations. In order to
get a more complete characterization of the entanglement, we also
make a quantum state tomography \cite{white1999, James2001} for our
narrow-band entangled source, the result is shown in
FIG.\ref{fig2}(b). From the tomography result, the calculated
fidelity between our state and $|\phi^-\rangle$ is $94.3\%$.

\begin{figure}[hbtp]
\includegraphics[width=\columnwidth]{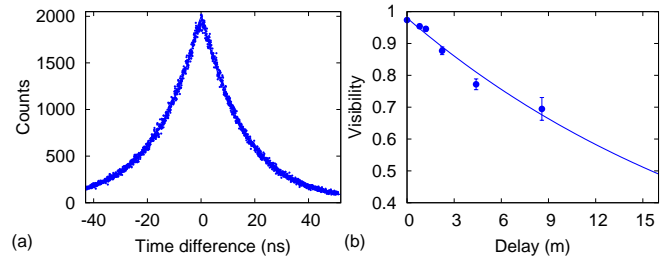}
\caption{(color online). (a) Time correlation measurement. The data
is fitted with a function of $c_0e^{-2\pi \Delta \nu |t|}$ The FWHM
correlation time is 23.0 ns. (b) Visibility in $|+\rangle /
|-\rangle$ basis as a function of the relative delay.The data is
fitted with a function of $v_0e^{-x/x0}$. Error bars represent
statistical errors.}\label{fig3}
\end{figure}

The correlation time between the down-converted photons is inversely
proportional to the bandwidth for a SPDC source \cite{Hong1985}.
Therefore, a narrow-band source at MHz level should exhibit a
correlation time which is much longer compared with the broad-band
SPDC source (typically on the order of several hundred fs). In our
experiment the detector signal of photon A is sent to a
time-to-amplitude converter (TAC) as the start signal, and the
signal of photon B is used as the stop signal. The TAC output signal
is sent to a multi-channel analyzer. The measured result is shown in
FIG. \ref{fig3}(a). The data is well agreed with the theoretical
expectation with the shape of $e^{-2\pi \Delta \nu |t|}$
\cite{Lu2000}. The best fit shows that the linewidth ($\Delta \nu$)
is about 9.6 MHz, which is well within the cavity linewidth. The
resolution time of the TAC utilized is about 50 ps, but the
single-photon detectors only have a resolution time of 350 ps, which
is comparable to the cavity round trip time of 520 ps. Therefore
\cite{Goto2003}, this time correlation measurement can not
distinguish our source from previous multi-mode cavity-enhanced SPDC
sources. To prove the single-mode property of our source we measure
the coherence length directly through the two-photon interference
experiment. This is done by observing the polarization correlation
visibility in $|+\rangle / |-\rangle$ basis between photon A and B,
as a function of the relative delay before PBS2. The result is shown
in FIG. \ref{fig3}(b). We find the coherence length to be $32\pm3$
m, which is consistent with the result from the time correlation
measurement in the relation of $\Delta L =
\frac{\nu}{\Delta\nu}\lambda$. While for a multi-mode source,
determined by the phase-matching condition, the coherence length is
usually less than several mm.

\begin{figure}[hbtp]
\includegraphics[width=.6\columnwidth]{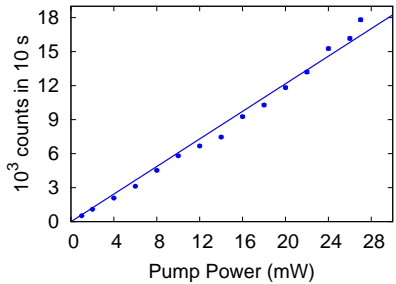}
\caption{(color online). Pair generation rate as a function of the
UV pump power.}\label{countpower}
\end{figure}

For a cavity-enhanced SPDC source, in the case of far below
threshold (about 1.88 W for our case), the pair generation rate is
proportional to the pump power, which is confirmed by our measured
result, which is shown in FIG. \ref{countpower}. At the pump power
of 27 mW, we get a maximal pair generation rate of 1780
$\rm{s}^{-1}$ and a corresponding maximal spectrum brightness of 185
$\rm{s}^{-1}\rm{MHz}^{-1}$. We fit the data with a proportional
function and find that the normalized spectrum brightness is about 6
$\rm{s}^{-1}\rm{MHz}^{-1}\rm{mW}^{-1}$. Further improvement of the
brightness could be possible by employing tighter focus, longer
crystal and a cavity with higher finesse.

In conclusion, we have experimentally generated narrow-band
polarization-entangled photon pairs through cavity-enhanced SPDC
with a linewidth of 9.6 MHz, which is comparable to the typical
linewidth of atomic quantum memories. Single-mode output is realized
by setting a reasonable cavity length difference between different
polarizations, using temperature controlled etalons and actively
stabilizing the cavity. The wavelength chosen is near Rubidium D2
line, making the storage a straightforward task using the method of
electromagnetically induced transparency (EIT) in cold atomic
ensembles \cite{Choi2008} if the UV pump is set to be pulsed. Since
the source is probabilistic and the entanglement is generated
through post-selection, one may think that it will limit its
applications in LDQC. According to a recent theoretical study
\cite{Zhao2007}, both of these problems could be eliminated if
applying the same trick. When combined with quantum memories, this
narrow-band entangled can be used to efficiently build entanglement
over large distance for LDQC, and to efficiently generate large
cluster states for LOQC, thus it will have extensive applications in
future scalable quantum information processing.

\begin{acknowledgments}
Thank C.-Z. Peng, B. Zhao and Y.-A. Chen for discussions. This work
is supported by the NNSF of China, the CAS, the National Fundamental
Research Program (under Grant No. 2006CB921900), and the ERC Grant.
\end{acknowledgments}

\bibliographystyle{my4author1}
\bibliography{myref}

\end{document}